\documentclass[aps,pre,floatfix,twocolumn,nofootinbib,showpacs]{revtex4} 
\usepackage{amsmath}
\usepackage{amssymb}
\usepackage{graphicx}
\usepackage{dcolumn}
\usepackage[sort&compress]{natbib}
\usepackage{bm}
\usepackage[dvips]{epsfig}
\newcommand{\comment}[1]{}

\newcommand{\CF}{{\cal F}}

\newcommand{\BEQ}{\begin{eqnarray}} 
\newcommand{\EEQ}{\end{eqnarray}} 
\newcommand{\BEA}{\begin{eqnarray}} 
\newcommand{\EEA}{\end{eqnarray}} 
 
\renewcommand{\d}{{\rm d}}

\newcommand{\E}{{\cal E}}

\newcommand{\half}{\frac{1}{2}} 
\begin{document} 

\title
{Non-equilibrium quantum fluctuations of work}
\author{A.E. Allahverdyan}
\affiliation{Yerevan Physics Institute,
Alikhanian Brothers street 2, Yerevan 375036, Armenia}

\begin{abstract} 
  The concept of work is basic for statistical thermodynamics. To gain
  a fuller understanding of work and its (quantum) features, it needs
  to be represented as an average of a fluctuating quantity. Here I
  focus on the work done between two moments of time for a thermally
  isolated quantum system driven by a time-dependent Hamiltonian. I
  formulate two natural conditions needed for the fluctuating work to
  be physically meaningful for a system that starts its evolution from
  a non-equilibrium state. The existing definitions do not satisfy
  these conditions due to issues that are traced back to
  non-commutativity. I propose a definition of fluctuating work that
  is free of previous drawbacks and that applies for a wide class of
  non-equilibrium initial states. It allows to deduce a generalized
  work-fluctuation theorem that applies for an arbitrary (out of
  equilibrium) initial state. \end{abstract}

\pacs{PACS: 05.30.-d, 05.70.Ln}


\maketitle

\section{ Introduction}

The first and second laws of statistical
thermodynamics are formulated using the concept of work, i.e. the
(average) energy exchanged by a system driven via a time-dependent
Hamiltonian \cite{balian,lindblad}. In this sense the work is a basic
quantity for thermodynamics. It is well-defined both in and out of
equilibrium for any (quantum or classical) system interacting with
external macroscopic work sources \cite{lindblad}.

However, the work as it appears in the first and second law is an
averaged quantity. There are at least two reasons why it is useful to
``de-average'' it, i.e. to present it as a random quantity. First, its
features are understood better in this way. Recall in this context
that the conservation of average energy for an isolated quantum system
is just a consequence of conserving energy eigenvalues and their
probabilities. Second, the current understanding of the second law is
that it has a statistical character and emerges out of averaging over
fluctuations \cite{demon,bassett}. Hence it is necessary to define
fluctuations of work for understanding e.g. the Thomson's formulation
of the second law \cite{bassett,lenard,thirring,bkphysica}. Both these
points are illustrated by fluctuation theorems; see
\cite{bkphysica,jar,maes,campisi_review,esposito_review} for reviews.

The existing definitions of quantum fluctuations of work can be
divided into 2 groups. Time-global definitions look for the work done
between two moments of time, as usual for any transfer quantity 
\cite{campisi_review,esposito_review,yukawa,tasaki,kurchan,
mukamel,lutz,sasaki,bkj,engel,gelin_kosov,campisi}. 
Time-local approaches adapt the
global definitions infinitesimally along an effective quantum
trajectory \cite{pekola,averin,maes_w,
chernyak,hu,wang,napoli_1,napoli_2,horowitz,campisi_th,fei}.

Here I focus on the time-global approaches (admittedly they are more
fundamental in the quantum case) for a thermally isolated dynamics and
note that they do not apply whenever the initial density matrix does
not commute with the (initial) Hamiltonian. This limitation is
essential, since work-extraction from non-equilibrium
(e.g. non-diagonal) states is important both conceptually
\cite{lenard} and practically \cite{abn}.

The aim of this paper is to present a definition of quantum
fluctuations of work that is free of the previous drawbacks. It is
based on the Terletsky-Margenau-Hill distribution
\cite{kirk,barut,marg,page,diosi,hartle}. The definition applies for a
class of initial density matrices that do not commute with the
(time-dependent) Hamiltonian.  It leads to a generalized fluctuation
theorem.

However, this definition is neither unique (otherwise there would not
be the issue with non-commutativity), nor it applies for an {\it
  arbitrary} initial state, because there it leads to negative
probabilities whose physical meaning is not clear. In this context, I
formulate 2 conditions for fluctuating work that are closely linked to
its physical meaning as the amount of energy exchanged with the source
of work. They need to be satisfied for any definition of the
fluctuating work and they hold for the presented one. It remains to be
seen whether this is indeed the most convenient definition or there
are even better ones to be uncovered in future.
\footnote{To explain why I decided to focus on the concept of work, I shall
compare its features to those of entropy production (EP). For a system
coupled to thermal baths, EP amounts to entropy increase of baths
\cite{balian,lindblad}. This definition does not apply more
generally|for non-equilibrium baths or thermally isolated case|since
the very definition of entropy is ambiguous there. For those cases, EP
is defined as an effective measure of irreversibility that has to be
positive and share the heuristics of entropy increase
\cite{kawai,hatano,ge_qian}. There is some consensus on how to define
EP for classical \cite{kawai,seifert} and semi-classical systems
\cite{esposito_review}.  But the quantum situation is ambiguous in
this respect \cite{roeck,deffner_epl}; e.g. Ref.~\cite{deffner_epl}
shows that there is a family of EPs associated with different notions
of effective phase-space. They lead to different expressions of the
(average) EP even for the initially equilibrium (Gibbsian) initial
state \cite{deffner_p}. These features differ from those of the
(average) work, which is well-defined for arbitrary (initial) states.}

This paper is organized as follows. Section \ref{setup} defines the
system to be studied. The next section reviews previous appeoaches and
explains why specifically they are not applicable out of
equilibrium. Section \ref{general} proposes two general conditions to
be satisfied for any definition of quantum fluctuating work. Section
discusses a new definition of fluctuating work that is free of
previous drawbacks. A generalized fluctuation theorem is derived and
interpreted in section \ref{generalized}. Section \ref{negative}
discusses certain limitations of the proposed approach. I summarize in
the last section. There are two Appendices.

\section{Set-up} 
\label{setup}

Consider a quantum system with an initial state
described by a density matrix $\rho$. The system is thermally
isolated: its dynamics is described by a time-dependent, Schroedinger
representation Hamiltonian $H(t)$ that generates a unitary evolution 
operator:
\BEA
 U_\tau=\overleftarrow{\exp} \left[ -\frac{i}{\hbar}
  \int_{0}^{\tau} \d
 t\,
 H(t) \right],
\label{uni:}
\EEA 
between the initial time $0$ and the final time $\tau$.
Here $\overleftarrow{\exp}$ denotes time-ordered exponent.
The (average) work $W$ done on the system reads
\cite{balian,lindblad}
\begin{eqnarray}
\label{work:} 
W&=&{\rm tr} [ \rho (H_{\rm F} -  H_{\rm I} )], \\
&&H_{\rm I} \equiv H(0), \qquad H_{\rm F}=U_\tau^\dagger H(\tau)U_\tau, 
\label{hei:}  
\end{eqnarray}
where $H_{\rm F}$ ($H_{\rm I}$) is the final (initial) Hamiltonian in
the Heisenberg representation. The definition of work applies to any
initial state: it is the average energy given up by the source of work
\cite{balian,lindblad}. Due to conservation of energy during the
system-work-source interaction, $W$ is the average energy transferred
to the source of work \cite{balian,lindblad}. This is seen explicitly
in approaches that deal with system-work-source interaction from the
first principles; see, e.g. \cite{picard,gemmer_mahler}. The intuitive
meaning of $W$ is that it is a ``high-graded'', mechanical energy that
can be wholly transferred from one work-source to another and
dissipated into heat.

$W$ can be observed in several ways, e.g. via the energy of the
work-source or by measuring the Heisenberg operator $H_{\rm F} -
H_{\rm I}$ at the final time.  Another (more usual) way of observing
$W$ is to consider an ensemble of identically prepared systems
(described by $\rho$) and divide it into 2 (equal) parts. Measuring
$H_{\rm I}$ ($H_{\rm F}$) on the first (second) part one recovers
${\rm tr}(\rho H_{\rm I})$ (${\rm tr}(\rho H_{\rm F})$); see
(\ref{work:}).  Thus, $W$ is directly observable and manifests the
energy conservation (first law) for the present problem.

Thus, I take the above definition of the average work $W$ as the basic
entity from which the fluctuating work is to be deduced under certain
additional assumptions.

Note that formally the above thermally isolated set-up applies also
for an open quantum system interacting with an environment
(e.g. thermal baths). Since the work is the energy transferred to the
source, one just needs to include the whole environment into a single
system interacting with the source. This is however a formal
procedure, because the environment is normally large and out of
control. Thus further research is needed to extend this set-up to open
systems. In this paper I focus on the thermally isolated set-up, also
because this is the first step towards understanding the more general
(open-system) situation.

\section{Two approaches for defining fluctuations of work}
\label{previous}

I now concentrate on two major (and different \cite{bkj,engel})
approaches for defining quantum fluctuations of work. My aim is to
compare these definitions to each other and to (\ref{work:}) and
understand where specifically they flaw in describing the fluctuating
work.

\subsection{Operator of work} 
\label{heiso}

The spirit of the Heisenberg representation is that time-dependent
operators are analogues of classical, time-dependent random
variables. Then the Heisenberg operator $H_{\rm F}-H_{\rm I}$ is
postulated to be the ``observable of work'' in the standard sense:
\cite{bkphysica,lindblad,yukawa,sasaki,bkj,gelin_kosov,chernyak,wang}
\cite{foo2} its eigenvalues are realizations of work and its
eigenvectors define the respective probabilities. I stress that (at
least formally) only one measurement (that of $H_{\rm F}-H_{\rm I}$)
is needed to obtaion the statistics of work according to this
definition. 

Now assume that the Schroedinger representation Hamiltonian changes
cyclically:
\begin{eqnarray}
  \label{eq:14}
H_{\rm I}=H(0)=H(\tau).    
\end{eqnarray}
One interpretation of (\ref{eq:14}) is that the system interacts with
the source of work only for $0\leq t\leq \tau$, i.e. it is strictly
isolated for $t<0$ and $t>\tau$: $H(t<0)=H(t>\tau)=H_{\rm I}$. 

Now since $H_{\rm I}$ and $H_{\rm F}$ have the same eigenvalues,
$H_{\rm F}-H_{\rm I}$ has eigenvalues of both signs. Since the approach
should apply for non-equilibrium initial states, we choose $H_{\rm I}$
and $H_{\rm F}$ such that $H_{\rm F}-H_{\rm I}$ has an eigenvalue
equal to zero. The corresponding eigenvector $|0\rangle$,
\BEA
\label{tut:}
(H_{\rm F}- H_{\rm I})|0\rangle=0, \EEA 
is taken as the initial state
$|0\rangle\langle 0|$. Due to
\begin{eqnarray}
  \label{eq:27}
[H_{\rm F},H_{\rm I}]\equiv H_{\rm
  F}H_{\rm I} - H_{\rm I}H_{\rm F}\not=0, 
\end{eqnarray}
$|0\rangle$ is neither an eigenstate of $H_{\rm F}$ nor an eigenstate
of $H_{\rm I}$. Eq.~(\ref{tut:}) implies that $H_{\rm F}- H_{\rm I}$
has on the state $|0\rangle\langle 0|$ a definite value equal to zero:
for {\it all single systems} from the ensemble described by
$|0\rangle\langle 0|$ no work is done and hence no energy is supposed
to be exchanged. But there are examples \cite{bkj}, showing that
(\ref{tut:}, \ref{eq:27}) are compatible with
\begin{eqnarray}
  \label{eq:4}
\langle 0| H_{\rm F}^m | 0\rangle=
\langle 0|U^\dagger_\tau H_{\rm I}^m U_\tau | 0\rangle
\not = \langle 0| H_{\rm I}^m | 0\rangle ~~{\rm for}~~ m>2.   
\end{eqnarray}
For a system that is strictly isolated for $t<0$ and $t>\tau$ [recall
(\ref{eq:14})], the inequality (\ref{eq:4}) implies that the
probabilities of some energies (i.e. the eigenvalues of $H_{\rm I}$)
do change due to the interaction with the source of work.

Thus, according to this definition it is possible to have energy
exchange with strictly zero fluctuations of work. In other words, the
link between energy exchange and the work done on a thermally isolated
system is generally absent. 

Though I tuned $H_{\rm F}- H_{\rm I}$ to have a zero eigenvalue, it is
clear that the problem is more general, e.g. it persists for $H_{\rm
  F}- H_{\rm I}$ having an eigenvalue close to zero \cite{foot}. 

I opine that due to this problem $H_{\rm F}- H_{\rm I}$ cannot be
interpreted as the work operator of for all initial states. Such an
interpretation can be perhaps kept for initial states $\rho$ that
commute with $H_{\rm F}$ or with $H_{\rm I}$ \cite{bkj}, but it is not
clear how to generalize this class of initial states.

\subsection{Two-time measurements of energy} 
\label{two-time}

We turn to the second approach
\cite{kurchan,tasaki,mukamel,campisi_review,esposito_review,lutz}.
Let the eigenresolution of the Schroedinger-representation Hamiltonian
$H(t)$ be
\begin{gather}
  \label{eq:3}
H(t)=\sum_k\, \epsilon_k(t)\,\E_k(t), \\ 
\E_k(t)\E_l(t)=\delta_{kl}\E_k(t), ~~~ {\rm
  tr}\,\E_k(t)={\rm const},\
\end{gather}
where $\epsilon_k(t)$ are the eigenvalues of $H(t)$, $\delta_{kl}$ is
the Kronecker symbol, and $\E_k(t)$ are the projector to the
corresponding eigen-space, whose dimension
${\rm tr}\,\E_k(t)$ is taken time-independent for simplicity.

Measuring $H(0)$ at $t=0$ produces 
$\epsilon_k (0)$ with probability
${\rm tr}(\rho\, \E_k(0))$ \cite{kurchan,tasaki,mukamel}. The
post-measurement state has the von Neumann-Luders form
$\rho_k\equiv\E_k(0)\rho\,\E_k(0)/{\rm tr}(\rho \E_k(0))$; it is then
evolved via (\ref{uni:}). At the final moment $\tau$ one measures
$H(\tau)$ and gets $\epsilon_l (\tau)$ with probability ${\rm
  tr}(U_\tau\rho_kU^\dagger_\tau\E_l(\tau))$, which is conditional
over the result $k$ of the first measurement.

The fluctuating work is presented as a {\it classical random variable}
with, respectively, realizations and probabilities\footnote{While
  this approach is standardly presented via two sharp measurements of
  energy, one can naturally wonder whether the same statistics of work
  can be approached via more feasible measurements; see
  \cite{watanabe} for a recent review of this issue.}
\begin{eqnarray}
  \label{eq:1}
&&  \epsilon_l (\tau)-\epsilon_k(0), \\ 
\widetilde{p}_{kl}&=& 
{\rm tr}\left(\,
\rho\, \E_k(0))\,{\rm tr}(U_\tau\rho_kU^\dagger_\tau\E_l(\tau)\,\right)\\
&=& {\rm tr}\left(\, \E_k(0)\,\rho\,\E_k(0)  
\,U^\dagger_\tau\E_l(\tau) U_\tau\,\right).
  \label{eq:11}
\end{eqnarray}

The problem of this definition is that it does not apply to initial
states that do not commute with $H(0)$: the average ``work''
$\widetilde{W}$ reads from (\ref{eq:1}, \ref{eq:11})
\begin{eqnarray}
  \label{eq:2}
  \widetilde{W}=\sum_{kl}\widetilde{p}_{kl}(\,\epsilon_l
  (\tau)-\epsilon_k(0)\,) 
  ={\rm tr}(\,\widetilde{\rho}\, (H_{\rm F}- H_{\rm I})\,), 
\end{eqnarray}
where 
\begin{eqnarray}
  \label{eq:6}
\widetilde{\rho} \equiv\sum_k \E_k(0)\,\rho\, \E_k(0).  
\end{eqnarray}
We obtain from (\ref{work:}, \ref{eq:2}, \ref{eq:6}):
\begin{eqnarray}
  \label{eq:19}
  W-\widetilde{W}&=&{\rm tr}(\,(\rho-\widetilde{\rho}\,) H_{\rm F}\,)
\nonumber\\
&=&\sum_{k\not=l} {\rm tr}(\,\E_k(0)\,\rho\, \E_l(0)\, H_{\rm F}\,).  
\end{eqnarray}
Hence for 
\begin{eqnarray}
  \label{eq:25}
  [\rho,H_{\rm I}]\not=0 ~~~ {\rm and}~~~[H_{\rm F},H_{\rm I}]\not=0,
\end{eqnarray}
(\ref{eq:1}, \ref{eq:11}) cannot be related to the work done on the
system with initial state $\rho$ by the external source, because
$\widetilde{W}\not=W$.  The physical reason for this conclusion is
that under $[\rho,H_{\rm I}]=0$ the first energy measurement (at
$t=0$) can be said to reveal the pre-existing (but unknown) value of
energy. In particular, the post-measurement density matrix does not
change: $\rho=\widetilde{\rho}$ [see (\ref{eq:6})]. In contrast, for
$[\rho,H_{\rm I}]\not =0$ already the first measurement is {\it
  invasive}: it leads to an irrversible change
$\rho\to\widetilde{\rho}$ of the density matrix that alters its
subsequent interaction with the source of work provided that $[H_{\rm
  F},H_{\rm I}]\not=0$; see Appendix A for a physical example. Put
differently, the reason for inapplicaility of the two-time measurement
approach is that it essentially alters the (non-equilibrium) initial
state \footnote{This point of altering the pre-measured state also
  appears in Ref.~\cite{kim}, where the authors study the energy
  changes for a system that couples to an external measuring apparatus
  and is thereby subject to projective measurements of a quantity that
  does not commute with energy (no work-source is supposed to be
  present). It is expected that in this situation the energy changes
  of the system will consist of both work and heat; no analysis of
  this problem is carried out in Ref.~\cite{kim}. }.

\subsection{Comparing two definitions with each other}
\label{compa}

We saw that the definition based on the operator of work does always
reproduce the average work (\ref{work:}), but it does not account
properly the notion of ``work = exchanged energy'' at least for some
initial states. I stress that this definition implies a one-time
approach, since one needs to measure the Heisenber operator $H_{\rm F}
- H_{\rm I}$ at the final time $\tau$.

The definition based on the two-time measurements of energy does not
reproduce the average work (\ref{work:}) if (\ref{eq:25}) holds. 

It is to be stressed that the drawbacks of both approaches do not show
up for $[\rho,H_{\rm I}]=0$. Hence if one is restricted by such
initial states, both approaches perform well, and it is a matter of
taste which one to prefer \footnote{When discussing this issue with
  people I met several times a viewpoint that Ref.~\cite{lutz} has
  shown that the quantum fluctuating work is not an operator,
  i.e. this reference ruled out the first definition. This is not
  correct: Ref.~\cite{lutz} shows that the work obtained via two-time
  measurements of energy cannot be (in general) represented as an
  outcome of an operator. But it does not point out any drawback of
  the Heisenberg-operator based definition, far from ruling out all
  possible definitions of the fluctuating work as an operator. }. 

Even then the operator definition has an advantage of being
time-symmetric: in contrast to the two-time energy measurement
approach, it applies not only for $[\rho,H_{\rm I}]=0$, but also for
$[\rho,H_{\rm F}]=0$ (and (\ref{eq:25})); see (\ref{hei:}, \ref{tut:},
\ref{eq:4}). 

\section{ General conditions}  
\label{general}

The above analysis of the two approaches
leads to the following general conditions demanded for the proper
definition of fluctuations of work.

{\it (i)} For cyclic changes of the Hamiltonian (cf. the discussion
above (\ref{tut:}, \ref{eq:4})), the zero fluctuations of work should
mean no energy exchange:
\begin{eqnarray}
  \label{eq:88}
  {\rm tr}(\rho H_{\rm I}^m)= {\rm tr}(\rho H_{\rm F}^m)\equiv 
  {\rm tr}(U_\tau\rho\,U^\dagger_\tau\,  H_{\rm I}^m)
  ~~ {\rm for}~~ m\geq 1.
\end{eqnarray}

{\it (ii)} The definition should apply for a possibly wide class of
initial states (including initial states that do not commute with the
initial Hamiltonian $H_{\rm I}$) and it should reproduce the average work
(\ref{work:}) for all initial states, where it applies.

As seen above, the first (second) condition does not hold for the
first (second) definition of fluctuations.

\section{Another definition for fluctuating work} 
\label{another}

\subsection{Estimation of energies via one measurement}

Below I work out a definition that satisfies the above two conditions,
and, similarly to (\ref{eq:1}), it presents the work as a classical
random quantity.  When discussing the approach based on two
measurements, we noted that its drawback stems from the invasive
character of the first measurement. It is then natural to {\it
  illustrate} a more general approach by avoiding the explicit
introduction of the first measurement. Hence at the final time $\tau$
we measure [cf. (\ref{hei:}, \ref{eq:3})]
\begin{eqnarray}
  \label{eq:20}
  H_{\rm F}=\sum_l \epsilon_l(\tau)\Pi_l, ~~ \Pi_l\Pi_{l'}=\Pi_l\delta_{ll'}.
\end{eqnarray}
Given the outcomes of this measurement, and provided that we know
$\rho$ and $H_{\rm I}$, we follow the ideas of \cite{luo,hall} and
introduce an approximation $f(H_{\rm F})$ of $H_{\rm I}$. The unknown
function $f(.)$ is sought from minimizing the mean-squared
difference (the simplest measure of magnitude):
\begin{eqnarray}
  \label{eq:17}
  {\rm tr}\left(
\, \rho\, (f(H_{\rm F})-H_{\rm I})^2
\right).
\end{eqnarray}
The minimization is straightforward [cf. (\ref{eq:20})]
 \cite{luo,hall}
\begin{eqnarray}
  \label{eq:21}
  f(H_{\rm F})=\sum_l f_l\Pi_l, ~~f_l\equiv\frac{{\rm Re}\, {\rm tr}
(\Pi_l\rho H_{\rm I}) }{{\rm tr} (\Pi_l\rho )}.
\end{eqnarray}
Now $f_l$ is represented via eigenvalues $\epsilon_k(0)$ of $H_{\rm I}$:
\begin{eqnarray}
  \label{eq:22}
&&  f_l=\sum_k\epsilon_k(0)p_{k|l}, ~~ p_{k|l}\equiv 
\frac{{\rm Re}\, {\rm tr}
(\Pi_l\E_k\rho) }{{\rm tr} (\Pi_l\rho )},
\end{eqnarray}
where $\E_k\equiv\E_k(0)$.
Provided that $p_{k|l}\geq 0$, it can interpreted as a conditional
probability for the initial energy to be $\epsilon_k(0)$. 
This condition does not hold automatically, but rather defines the
class of states and Hamiltonians for which it is legitimate to
interpret $p_{k|l}$ as probabilities; see section \ref{negative}.

\subsection{Definition of fluctuating work}

Hence fluctuating work is defined as a classical random quantity with,
respectively, realizations and probabilities:
\begin{eqnarray}
  \label{eq:5}
  \epsilon_l (\tau)-\epsilon_k(0), ~~~~ 
p_{kl}\equiv{\rm Re} \, {\rm tr}(\rho
  \E_k\Pi_l)\geq 0.
\end{eqnarray}
Now $p_{kl}\geq 0$ is interpreted as the joint probability for the
eigenvalues of $H_{\rm I}$ and $H_{\rm F}$. 
We could avoid the reasoning of (\ref{eq:17}--\ref{eq:22}) and just
introduce (\ref{eq:5}) as a postulate.

Note that $p_{kl}$ has correct marginal probabilities 
\begin{eqnarray}
\label{puk}
\sum_k p_{kl}={\rm tr}(\rho
\Pi_l), \qquad 
\sum_l p_{kl}={\rm tr}(\rho
\E_k).   
\end{eqnarray}

For $[\rho,H_{\rm I}]=0$ we revert from (\ref{eq:5}) to (\ref{eq:1},
\ref{eq:11}) using $\E_k\rho=\E_k\rho\E_k$.

The Cauchy-Schwartz inequality implies $p^2_{kl}\leq 1$:
\begin{eqnarray}
  [{\rm Re} \, {\rm tr}(\rho
   \E_k\Pi_l)]^2
\leq |{\rm tr}(\sqrt{\rho}
   \E_k\Pi_l\sqrt{\rho})|^2\leq 
{\rm tr}(\rho
 \Pi_l){\rm tr}(\rho
   \E_k)\leq 1.\nonumber 
\end{eqnarray}
But for specific choices of $\rho$, $p_{kl}$ can turn negative for
given $\E_k$ and $\Pi_l$, and then its interpretation as a joint
probability is lost; see section \ref{negative}. From now on and till
(\ref{eq:24}) we assume that $p_{kl}\geq 0$.

Condition {\it (ii)} holds, since the first and the second moment
calculated from (\ref{eq:5}) are equal, respectively, to the first and
second moments of the operator $H_{\rm F}-H_{\rm I}$:
\begin{gather}
  \label{eq:7}
  {\rm tr}(\rho (H_{\rm F}-H_{\rm I})^m)={\sum}_{k,l}\, p_{kl}
  (\epsilon_l (\tau)-\epsilon_k(0))^m, ~ m=1,2.
\end{gather}
However, already the third moments generally differ, the difference
involving a double-commutator [cf. (\ref{eq:20}, \ref{eq:5})]:
\begin{gather}
{\rm tr}(\rho (H_{\rm F}-H_{\rm
    I})^3) -   \label{eq:8} {\sum}_{k,l}\, p_{kl}
  (\epsilon_l (t)-\epsilon_k(0))^3
\nonumber\\
={\rm tr}(\rho [\,\frac{H_{\rm F}+H_{\rm I}}{2},[\,H_{\rm F},H_{\rm I}\,]\,]\,).
\end{gather}
Let us check that condition {\it (i)} holds. For cyclic, $H(t)=H(0)$,
change of the Hamiltonian, the zero fluctuations of work mean 
\begin{eqnarray}
  \label{eq:9}
  p_{kl}={\rm Re} \, {\rm tr}(\rho
  \E_k\Pi_l)=0~~ {\rm for}~~{\rm all}~~ k\not =l. 
\end{eqnarray}
Employing $0={\rm Re} \, {\rm tr}(\rho
  \E_k\Pi_l)={\rm Re} \, {\rm tr}(\rho
  (1-\sum_{k'\not=k}\E_{k'})\Pi_l)$ and 
re-arranging the terms we get
  \begin{eqnarray}
    \label{eq:10}
    {\rm tr}(\rho
 \E_l)
=   {\rm tr}(\rho
 \Pi_l)
=
   {\rm Re} \, {\rm tr}(\rho
  \E_l\Pi_l) ~~ {\rm for}~~{\rm all}~~ l.
\end{eqnarray}
The first equality here suffices to establish (\ref{eq:88}).  If
$\rho$ does not have zero eigenvalues one can find from (\ref{eq:9})
stronger conditions, but we shall not dwell on that.

In section \ref{compa} we noted that the two-time energy measurement
approach does not apply when $[\rho,H_{\rm F}]=0$ but (\ref{eq:25})
holds. It is now seen that the present definition does not have this
drawback: for $[\rho,H_{\rm F}]=0$ we obtain
\begin{eqnarray}
  \label{eq:16}
  p_{kl}= {\rm tr} (\E_k\Pi_l\rho\Pi_l) \geq 0. 
\end{eqnarray}
This expression is intuitive, but (for $[\rho,H_{\rm I}]\not=0$) it
cannot be obtained from the two-time approach, where one
first measures energy at $t=0$ and then at $t=\tau>0$.

\subsection{Discussion}

The joint probability $p_{kl}$ for non-commuting variables was
introduced in \cite{kirk,barut,marg} (Terletsky-Margenau-Hill
distribution). Though it is one of many possible definitions of joint
probabilities for non-commuting variables, it is very convenient in
the context of quantum statistical mechanics. This point was made in
\cite{barut,marg} and we shall confirm it below when deriving the
generalized fluctuation theorem.  As many other joint probabilities
(e.g. the Wigner function), $p_{kl}$ can be measured experimentally
\cite{bambo}. Note that $f_l$ in (\ref{eq:21}, \ref{eq:22})
corresponds to the generalized weak value \cite{agarwal}, which is
alternatively known as the locally averaged value of energy
\cite{cohen}; this interpretation was employed in (\ref{eq:17}
-\ref{eq:22} ).  The relation between $f_l$ and $p_{kl}$ was noted in
\cite{sokol,ozawa,sagawa}. Also, the form of $p_{kl}$ leads to the
most general consistency condition in the history approach to quantum
mechanics \cite{page,diosi,hartle} \cite{foo2}. $p_{kl}$ behaves
expectedly under coarse-graining: when two orthogonal subspaces
(e.g. described, respectively, by projectors $\Pi_1$ and $\Pi_2$) are
joined into one space (described by $\Pi_1+\Pi_2$), the probabilities
are added:
\begin{eqnarray}
  \label{co}
p_{k1}+p_{k2}={\rm Re}\, {\rm
 tr}\left(\,(\Pi_1+\Pi_2)\E_k\rho\,\right).  
\end{eqnarray}
Ref.~\cite{johansen} derives $p_{kl}$ axiomatically and underlines
another deep feature of $p_{kl}$: it is time-symmetric, i.e. invariant
with respect to interchanging $\Pi_l$ with $\E_k$. We already noted
this feature around (\ref{eq:16}).
 
\subsection{Summary}

Let us briefly summarize recalling why $p_{kl}$ defined in
(\ref{eq:5}) can be regarded as a joint distribution for initial and
final energies.

-- It emerges out of estimating two non-commuting observables via one
measurement; see (\ref{eq:20}--\ref{eq:22}) . 

-- Whenever any two among three operators $\Pi_l$, $\E_k$, $\rho$
commute, $p_{kl}$ reduces to the expected form ${\rm
  tr}[\,\Pi_l\E_k\rho \, ]$.

-- $p_{kl}$ has correct marginals; see (\ref{puk}).

-- It is time-symmetric and linear with respect to projectors $\Pi_l$
and $\E_k$. Hence $p_{kl}$ is additive, much in the same way as the
ordinary probability ${\rm tr}[\rho\E_k]$.

\section{Generalized fluctuation theorem} 
\label{generalized}

\begin{figure}[htbp]
\includegraphics[width=8cm]{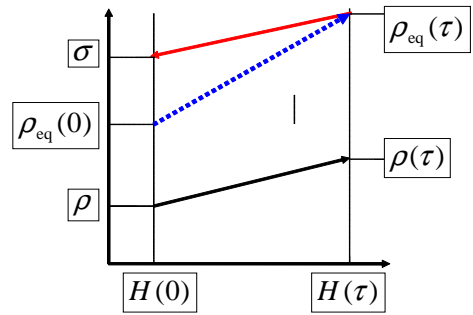}
\caption{ Three processes that appear in (\ref{eq:13}). They are
  depicted in a schematic coordinate plane with the x-axes (y-axes)
  being Hamiltonian (density matrices) in the Schroedinger
  representation. Black (lower, full) arrow: the target thermally
  isolated process. Blue (dashed) arrow: isothermal process. Red
  (upper, full) arrow: another thermally isolated process. Notations
  refer to (\ref{eq:3}, \ref{eq:13}, \ref{kora}).}
\label{fig_1}
\end{figure}

\subsection{Derivation and interpretation} 

Following the logic of the
equilibrium fluctuation theorem we take a parameter $\beta$ and work
out using (\ref{eq:5}):
\begin{eqnarray}
  \label{eq:12}
  \sum_{kl}p_{kl} e^{-\beta (\epsilon_l(\tau)-\epsilon_k(0)  )}={\rm Re}
  \sum_{kl}{\rm tr}(\rho \Pi_le^{-\beta \epsilon_l(\tau)}\E_k
  e^{\beta\epsilon_k(0)  }).
\nonumber
\end{eqnarray}
We get from this the following fluctuation theorem
\begin{eqnarray}
  \label{eq:13}
  \left\langle e^{-\beta (w-\Delta \CF)}
\right\rangle ={\rm Re}\, {\rm tr}\left( \sigma \,
\rho^{-1}_{\rm eq}(0) \, \rho \right)\equiv \Upsilon,
\end{eqnarray}
where 
$w=\epsilon_l(\tau)-\epsilon_k(0)$ are realizations of the random
work, $\langle ...\rangle$ means averaging over $p_{kl}$,
and where 
\begin{gather}
\label{kora}
\rho_{\rm eq}(t) \equiv e^{-\beta H(t) }\left / {\rm tr}(e^{-\beta
    H(t)})\right., \qquad  0\leq t \leq \tau, \\
\label{tora}
\beta\Delta\CF\equiv -\ln\,{\rm tr}[e^{-\beta
  H(t)}]
+\ln\,{\rm tr}[e^{-\beta H(0)}], \\
\label{mora}
\sigma\equiv U_\tau^\dagger \rho_{\rm eq}(\tau) U_\tau .
\end{gather}
Eq.~(\ref{eq:13}) relates to each other 3 processes; 
see Fig.~\ref{fig_1}. The first of them
is the thermally isolated process we focused on: the system starts
from the density matrix $\rho$ and Hamiltonian $H(0)$ and (in the
Schroedinger representation) ends at density matrix
$\rho(\tau)=U_\tau\rho U_\tau^\dagger$ and Hamiltonian $H(\tau)$. The
work $w$ and averaging $\langle ...\rangle$ in (\ref{eq:13}) refer to
this process.

For the second process we imagine that the system (at some pre-initial
time) is attached to a thermal bath at temperature $1/\beta$, and
relaxes to the Gibbsian equilibrium density matrix $\rho_{\rm eq}(0)$;
cf. (\ref{kora}). Then it follows an isothermal quasi-equilibrium
process, where the Hamiltonian slowly changes from $H(0)$ to
$H(\tau)$, under a weak but fixed coupling with the bath.  Since the
change is slow, the density matrix during the process equals
$\rho_{\rm eq}(t)$, $0\leq t\leq \tau$. The work done in this process
is given the equilibrium free energy difference $\Delta\CF$ in
(\ref{eq:13}, \ref{tora}) \cite{balian,lindblad}.

During the third process the system at the end of the previous
isothermal process is decoupled from the bath and undergoes the
reversal of the first thermally isolated process. The final
density matrix $\sigma$ of this process appears in
(\ref{eq:13}, \ref{mora}).

In the (initially) equilibrium situation $\rho_{\rm eq}(0)=\rho$, we
revert to the usual (equilibrium) fluctuation theorem 
\begin{eqnarray}
  \label{eq:18}
\left\langle e^{-\beta (w-\Delta \CF)} \right\rangle=1.  
\end{eqnarray}
This theorem relates together characteristics of the first (thermally
isolated) and second (isothermal) process. Note that the approach
based on two-time measurements of certain observables (not necessarily
energy) can also generate fluctuation theorems whose r.h.s. is not
equal to $1$ [an analogue of $\Upsilon$, cf. (\ref{eq:13})]
\cite{kafri_deffner,rastegin,albash}. There, however, the initial
(post-measurement) state always commutes with the first observable, in
contrast to (\ref{eq:13} ), which holds for an arbitrary initial
state.

For the equilibrium fluctuation theorem (\ref{eq:18}) we note that the
existence of the bath is necessary for defining the second process (at
least when the system is finite, as we assume here). Without the bath,
i.e. when the second process is also thermally isolated, the work
during the slowest, reversible process is generally {\it not} given by
the free energy difference \cite{minima}. There is a simple way to see
this fact explicitly: any unitary time-evolution conserves eigenvalues
of the density matrix:
\begin{eqnarray}
  \label{eq:23}
  {\rm Spectrum}[\,U_\tau \rho_{\rm
    eq}(0)U_\tau^\dagger\,] ={\rm Spectrum}[\,\rho_{\rm
    eq}(0)\,].  
\end{eqnarray}
Hence $\rho_{\rm eq}(\tau)$ in (\ref{kora}) cannot be
obtained from $\rho_{\rm eq}(0)$ via a unitary process. Put
differently, the equilibrium fluctuation theorem (\ref{eq:18}) does
not generally characterizes the amount of irreversibility (slow versus
fast realization) of the thermally isolated process.  Instead it
compares two different processes.

Finally, let us again look at (\ref{eq:13}) and compare it with
(\ref{eq:18}): the equilibrium fluctuation theorem (\ref{eq:18}) has
precisely the same form as the corresponding classical fluctuation
relation. This is related to the fact that the equilibrium initial
state $\rho_{\rm eq}(0)$ has classical features with respect to the
(initial) energy distribution \footnote{Quantum effects are carefully
  hidden under (\ref{eq:18}); see e.g. \cite{vedved}.}.  In contrast,
(\ref{eq:13}) retains quantum features, since its right-hand-side
contains non-commutative quantities.

\subsection{Work-free-energy relation}

 Using convexity,
 $\langle e^x\rangle \geq e^{\langle x\rangle}$, we deduce
from (\ref{eq:5}, \ref{eq:13}) a generalization of the usual work-free-energy
relation:
\begin{eqnarray}
  \label{eq:15}
-\beta (W-\Delta\CF)\geq \ln \Upsilon\equiv 
\ln\left(\sum_{kl}\frac{\mu_l}{\nu_k}p_{kl}\right),  
\end{eqnarray}
where $\mu_l$ and $\nu_k$ are the eigenvalues of $\sigma$ and
$\rho_{\rm eq}(0)$, respectively. They directly relate to eigenvalues
of $H(\tau)$ and $H(0)$. Now $\ln\Upsilon$ can be arbitrary large,
e.g. when one of $\nu_k$ is close to zero. Then the equilibrium
relation 
\begin{eqnarray}
\label{lola}
-\beta (W-\Delta\CF)\geq 0,  
\end{eqnarray}
carries out to non-equilibrium. Note that $\ln\Upsilon\geq 0$ (which
guarantees (\ref{lola})) is not always true; see Appendix B.

\section{ Negativity of $p_{kl}$}  
\label{negative}

The above theory for fluctuations of work was developed under
assumption $p_{kl}\geq 0$ (cf. (\ref{eq:5} )) (though we shall see
that formally not all results demand this assumption).
However, for given projectors $\E_k$ and $\Pi_l$
with $[\E_k,\Pi_l]\not =0$, there are $\rho$'s such that
\begin{eqnarray}
  \label{eq:24}
 p_{kl}= {\rm
tr}(\rho X_{kl})<0, 
~~ X_{kl}\equiv\half(\E_k\Pi_l+\Pi_l\E_k).
\end{eqnarray}
This is because for $[\E_k,\Pi_l]\not =0$, $X_{kl}$ has at least one
negative eigenvalue \cite{hartle}, e.g. for one-dimensional projectors
$\E_k$ and $\Pi_l$ the non-zero eigenvalues of $X_{kl}$ are 
\begin{eqnarray}
  \label{eq:29}
\half
({\rm tr}(\E_k\Pi_l)\pm \sqrt{{\rm tr}(\E_k\Pi_l)}).  
\end{eqnarray}
More generally,
for $[\E_k,\Pi_l]\not =0$ there is a vector $|\psi\rangle$ so that 
$\E_k|\psi\rangle=0$, but $\E_k\Pi_l|\psi\rangle\not=0$. Let now 
$X_{kl}=\sum_a x_a |x_a\rangle\langle x_a|$ be the eigen-resolution of
$X_{kl}$, and $x_1$ be the smallest eigen-value of $X_{kl}$.
We have
\begin{eqnarray}
  \label{eq:26}
  x_1\leq\sum_a x_a|\langle x_a|\psi\rangle|^2={\rm Re}\langle
  \psi|\E_k\Pi_l|\psi\rangle
=0.
\end{eqnarray}
This proves that at least the smallest eigenvalue of $X_{kl}$ is
negative, since for $[\E_k,\Pi_l]\not =0$ the inequality in
(\ref{eq:26}) is strict. (It turns into equality for $[\E_k,\Pi_l]=0$,
in which case $x_1=0$.)  The magnitude of this negativity can be
estimated from \footnote{Inequalities in (\ref{eq:28}) wede derived in
  \cite{zhang} for a slightly more general case of two non-negative
  operators (not necessarily projections). The first [second]
  inequality follows from $(\E_k+\Pi_l-\frac{1}{2})^2\leq 1$
  [$(\E_k-\Pi_l)^2\geq 0$] using $\E_k=\E_k^2\leq 1$ and $\Pi_l=\Pi_l^2\leq
  1$. }
\begin{eqnarray}
  \label{eq:28}
  -\frac{1}{8}\leq X_{kl}\leq 1,
\end{eqnarray}
where $1$ is the unit operator, and e.g. $X_{kl}\leq 1$ means that the
eigenvalues of $1-X_{kl}$ are non-negative. Thus the smallest
eigenvalue of $X_{kl}$ is not smaller than $  -\frac{1}{8}$ (this is
consistent with (\ref{eq:29})).

Whenever $p_{kl}<0$, the usual probability interpretation for
$p_{kl}$|and hence the presented definition of fluctuations of work|do
not apply. Nevertheless, the expression (\ref{eq:7}) for the first and
second moments of work, as well as the fluctuation theorem
(\ref{eq:13}), still apply {\it formally}, i.e.  their derivations do
not require the validity of $p_{kl}\geq 0$. This condition is
demanded, e.g. for (\ref{eq:15}).

However, the positive eigenvalues of $X_{kl}$ are larger than negative
one(s), e.g. due to ${\rm tr}[\,X_{kl}]= {\rm tr}[\, \Pi_l \E_k
\Pi_l\,]\geq 0$. Also, in certain cases of $p_{kl}<0$ we can follow
the reasoning of (\ref{co}) and still define positive probabilities by
coarse-graining $p_{kl}$.

\section{Summary}

This paper is started by studying the applicability of the existing
definitions of fluctuating work to non-equilibrium initixal states of a
quantum system subject to a thermally siolated process. The approach
based on two-time energy measurements do not apply for initial states
that do not commute with the initial Hamiltonian, because it does not
properly reproduce the average work; see section \ref{two-time}. The
applicability domain of the operator definition of work is wider, but
it is still limited, because this definition does not support (for
non-equilibrium initial states) the relation between the work and
energy change; see section \ref{heiso}.

The route to defining quantum fluctuating work goes via formulating
necessary physical conditions which possible definition should hold. I
propose in section \ref{general} that there are (at least) two such
restrictions: the fluctuating work should relate to energy change and
it should respect the definition (\ref{work:}) of the average work.

I worked out in section \ref{another} (what seems to me) the simplest
definition of the fluctuating work that holds the above two
features. This definition does apply to class of non-equilibrium
initial states. Its applicability domain is clearly defined by the
non-negativity $p_{kl}\geq 0$ of joint probabilities; see (\ref{eq:5})
and section (\ref{negative}). 

This definition employes only one measurement (by analogy to
the definition of work based on the Heisenberg operator (\ref{hei:});
see section \ref{heiso}). For initial states that commute with the
initial Hamiltonian this definition reduces to what is obtained with
the two-time energy measurements. 

I believe that this definition of fluctuating work does advance our
understanding of non-equilibrium statistical mechanics, e.g. it allows
to derive a generalized fluctuation theorem, which connects together 3
related processes; see section \ref{generalized}.

\section*{Acknowledgements}

I acknowledge discussions with K. Hovhannisyan, E. Sherman,
D. Sokolovskii, M. Campisi and M. Perernau-Llobet. I was supported by
COST network MP1209.

\appendix

\section{Fluctuations of work for the Rabi's model}

\subsection{The Rabi's model}

The purpose of this section is to illustrate the definition of work
fluctuations (proposed in the main text) for the Rabi's model: a
two-level system driven by an oscillating external field
\cite{william,vedral}.  I also contrast with each other different
definitions of fluctuating work.

There are several reasons why I choose to illustrate the theory of
fluctuating work with this specific model.

-- The model is basic for several
fields (quantum optics, NMR/ESr physics {\it etc}). 

-- It is exactly solvable.

-- The non-equilibrium initial states for this model are theoretically
natural and experimentally realizable. 

The time-dependent Hamiltonian of the model reads \cite{william,vedral} 
\begin{eqnarray}
  \label{eq:70}
  H(t)=\frac{\omega}{2}\,\sigma_z+\frac{g}{2}[\sigma_x\, \cos(\omega
  t)+\sigma_y\, \sin(\omega t)],
\end{eqnarray}
where $\omega>0$ is the (free) frequency of the two-level system,
while $g$ quantifies the coupling with the external field. 

Here $\sigma_{x,y,z}$ are Pauli matrices. We shall write them in the
representation of up $|\uparrow\rangle$ and down $|\downarrow\rangle$
spin states:
\begin{gather}
  \label{eq:701}
  \sigma_z=|\uparrow\rangle\langle \uparrow| - 
|\downarrow\rangle\langle \downarrow|, ~~
  \sigma_x=|\uparrow\rangle\langle \downarrow| +
|\downarrow\rangle\langle \uparrow|,\\
  \sigma_y=-i(|\uparrow\rangle\langle \downarrow| -
|\downarrow\rangle\langle \uparrow|).
\end{gather}

Eq.~(\ref{eq:70}) assumes the resonant case, where the frequencies of
the two-level system and external are both equal $\omega$ (we take
$\hbar=1$) \cite{william,vedral}.

The eigenvalues 
\begin{eqnarray}
  \label{eq:71}
  \epsilon_1=\frac{1}{2}\sqrt{g^2+\omega^2}, ~~~
  \epsilon_2=-\frac{1}{2}\sqrt{g^2+\omega^2},
\end{eqnarray}
of $H(t)$ do not depend on time. The orthogonal and normalized
eigenvectors of $H(t)$ read ($| \epsilon_{1,2}(t)\rangle$ are
row-vectors)
\begin{eqnarray}
  \label{eq:72}
  \langle \epsilon_1(t)|
=\frac{
\left( (\omega+\sqrt{\omega^2+g^2})\, e^{i\omega t},\, g    \right)}
{\sqrt{g^2+(\omega+\sqrt{\omega^2+g^2\,\,})^2\,\,}}, \\
  \label{eq:73}
  \langle \epsilon_2(t)|
=\frac{
\left( (\omega-\sqrt{\omega^2+g^2})\, e^{i\omega t},\, g    \right)}
{\sqrt{g^2+(\omega-\sqrt{\omega^2+g^2\,\,})^2\,\,}}.
\end{eqnarray}
The unitary operator generated by (\ref{eq:70}) is
\cite{william,vedral}
\begin{eqnarray}
  \label{eq:74}
  U_t=\exp\left[  \frac{-i\omega t}{2}\,\sigma_z\right]
\exp\left[  \frac{-ig t}{2}\,\sigma_x\right].
\end{eqnarray}
It satisfies the evolution equation $i\partial_t U_t=H(t)U_t$, as
verified by direct substitution. 

\subsection{Fluctuations of work}

The advantage of this situation is that the up and down initial states
and their mixtures are natural initial states
\cite{william,vedral}. We thus take the initial state as
\begin{eqnarray}
  \label{eq:75}
\rho=\frac{1-\delta}{2}\,|\uparrow\rangle\langle \uparrow| +
\frac{1+\delta}{2}\,|\downarrow\rangle\langle \downarrow|, ~~
\end{eqnarray}
where $|\delta|\leq 1$ is a parameter. For applications in NMR/ESR
physics, $|\delta|$ is a small dimensionless number, e.g.
$|\delta|\sim 10^{-2}$ \cite{william}; it can be significantly larger
in quantum optics \cite{vedral}.

Clearly, the initial state $\rho$ does not commute with the initial
Hamiltonian $H(0)$ (excluding the case $\delta=0$) [cf. (\ref{eq:70},
\ref{eq:72}, \ref{eq:73})], so we are in the situation described in
the main text.

According to (16) of the main text, we get from (\ref{eq:71}) three
values of the the fluctuating work (\ref{eq:76}, \ref{eq:77},
\ref{eq:78}) with their respective probabilities (\ref{eq:76a},
\ref{eq:77a}, \ref{eq:78a}) (with an obvious adaptation of notations):
\begin{gather}
  \label{eq:76}
 \sqrt{g^2+\omega^2}, ~~ \\  
  \label{eq:76a}
 p_{+}={\rm Re}\left\{\,
\langle \epsilon_2(0)|U^\dagger_t|\epsilon_1(t)\rangle\,
\langle \epsilon_1(t)|U_t\rho|\epsilon_2(0)\rangle\,
\right\};
\end{gather}
\begin{gather}
  \label{eq:77}
 -\sqrt{g^2+\omega^2}, ~~ \\  
  \label{eq:77a}
 p_-={\rm Re}\left\{\,
\langle \epsilon_1(0)|U^\dagger_t|\epsilon_2(t)\rangle\,
\langle \epsilon_2(t)|U_t\rho|\epsilon_1(0)\rangle\,
\right\};
\end{gather}
\begin{gather}
  \label{eq:78}
 0, ~~ \\  
 p_0=1-p_+-p_- .
  \label{eq:78a}
\end{gather}

Eqs.~(\ref{eq:72}, \ref{eq:73}, \ref{eq:74}, \ref{eq:76a},
\ref{eq:77a}, \ref{eq:78a}) imply
\begin{eqnarray}
  \label{eq:79a}
&&  p_{+}=\frac{\omega^2 \sin^2[gt/2]}{2(g^2+\omega^2)}\,
\left(1+\delta\sqrt{\frac{g^2}{\omega^2}+1}\,\right),\\
\label{eq:79b}
&&  p_{-}=\frac{\omega^2 \sin^2[gt/2]}{2(g^2+\omega^2)}\,
\left(1-\delta\sqrt{\frac{g^2}{\omega^2}+1}\,\right),\\
&& p_{0}=\frac{g^2+\omega^2 \cos^2[gt/2]}{g^2+\omega^2}.
\end{eqnarray}
Note that $p_+-p_-$ and $\delta$ have the same sign, and this agrees
with the logics of the second law (even though the initial state is
not in equilibrium): $\delta>0$ means the lower (down) initial state
is more populated [cf. (\ref{eq:75})] and hence the probability of
energy increase is larger: $p_+>p_-$.

The average work agrees with (\ref{eq:79a}, \ref{eq:79b}):
\begin{eqnarray}
  W={\rm tr}\left(\, U_t\rho U_t^\dagger H(t)-\rho H(0)
    \,\right)&=&\sqrt{g^2+\omega^2} (p_+-p_-)\nonumber\\
             &=&\delta\omega\sin^2[gt/2].
  \label{eq:79c}
\end{eqnarray}

Eqs.~(\ref{eq:79a}, \ref{eq:79b}) are non-negative|and hence qualify as
probabilities|for
\begin{eqnarray}
  \label{eq:81}
  1\geq |\delta|\sqrt{\frac{g^2}{\omega^2}+1},
\end{eqnarray}
i.e. for a sufficiently mixed initial state ($|\delta|$ is not close
to $1$), and/or for a sufficiently small $\frac{g^2}{\omega^2}$
(relatively weak influence on the two-level system). Condition
(\ref{eq:81}) does not hold, and hence either $p_{12}$ or $p_{21}$ is
negative for $|\delta|=1$ (initially pure state).

\subsection{Two-time measurements of energy}

Now the two-time measurement approach produces the same 3 realizations
$(\pm\sqrt{g^2+\omega^2},\, 0)$, but their probabilities are different:
\begin{gather}
  \label{eq:776}
\sqrt{g^2+\omega^2}, \\  
  \label{eq:776a}
\widetilde{p}_{+}=\langle \epsilon_2(0)|\rho|\epsilon_2(0)\rangle\,
\langle \epsilon_2(0)|U^\dagger_t|\epsilon_1(t)\rangle\,
\langle \epsilon_1(t)|U_t|\epsilon_2(0)\rangle\,
\end{gather}

\begin{gather}
  \label{eq:777}
-\sqrt{g^2+\omega^2}, \\  
  \label{eq:777a}
\widetilde{p}_-=\langle \epsilon_1(0)|\rho|\epsilon_1(0)\rangle\,
\langle \epsilon_1(0)|U^\dagger_t|\epsilon_2(t)\rangle\,
\langle \epsilon_2(t)|U_t|\epsilon_1(0)\rangle\,
\end{gather}

\begin{gather}
0, \\  
  \label{eq:777aa}
1-\widetilde{p}_--\widetilde{p}_+ .
\end{gather}
The difference between (\ref{eq:76a}, \ref{eq:77a}) and
(\ref{eq:776a}, \ref{eq:777a}) is best visible without working out
(\ref{eq:776a}, \ref{eq:777a}) but looking directly to the average
produced by (\ref{eq:776a}, \ref{eq:777a}):
\begin{eqnarray}
  \label{eq:900}
\sqrt{g^2+\omega^2}\, (\widetilde{p}_+-\widetilde{p}_-)
             =\frac{\delta\omega^3\sin^2[gt/2]}{g^2+\omega^2}.
\end{eqnarray}
It is seen that (\ref{eq:900}) does differ from the average work
(\ref{eq:79c}), and hence the approach based on the two-time
measurements of energy does not apply.

\subsection{Operator of work}

Let us now turn to the operator of work approach. This operator is
given as
\begin{eqnarray}
  \label{eq:901}
  \Delta H(t)=U^\dagger_t H(t)U_t-H(0).
\end{eqnarray}
As follows from (\ref{eq:70}, \ref{eq:74}), $\Delta H(t)$
has eigenvalues and (respective) eigenvectors:
\begin{eqnarray}
  \label{eq:902a}
&& \Delta_1= w\sin[gt/2], \\ 
  \label{eq:902b}
&& \langle \Delta_1|=\frac{\left(\,- i(\,
\sin[gt/2]-1),\, \cos[gt/2] \, \right)}{\sqrt{2(1-\sin[gt/2])}}, \\
  \label{eq:903a}
&& \Delta_2= -w\sin[gt/2], \\ && \langle \Delta_2|=\frac{\left(\,- i(\,
\sin[gt/2]+1),\, \cos[gt/2] \, \right)}{\sqrt{2(1+\sin[gt/2])}}.
  \label{eq:903b}
\end{eqnarray}
On the initial state $\rho$ each of these eigenvalues is realized with
probabilities
\begin{eqnarray}
  \label{eq:904}
  \langle \Delta_1|\rho |\Delta_1\rangle 
=\frac{1+\delta \sin[gt/2]}{2}, \\
  \langle \Delta_2|\rho |\Delta_2\rangle 
=\frac{1-\delta \sin[gt/2]}{2}. 
  \label{eq:905}
\end{eqnarray}
Now the average work (\ref{eq:79c}) is expectedly reproduced from
(\ref{eq:901}--\ref{eq:905}): 
\begin{eqnarray}
W=\sum_{k=1,2}\Delta_k \langle
\Delta_k|\rho |\Delta_k\rangle 
\end{eqnarray}

We compare predictions of the operator of work approach with
(\ref{eq:76}--\ref{eq:78a}). According to (\ref{eq:76}, \ref{eq:77},
\ref{eq:78}) there are 3 time-independent realizations of work, while
in (\ref{eq:902a}, \ref{eq:903a}) there are 2 time-depedent
realizations $\pm \omega\sin[gt/2]$. Note that the eigenvalues of the
operator of work $\pm \omega\sin[gt/2]$ nullify simultaneously with
probabilities (\ref{eq:76a}, \ref{eq:77a}) for non-zero values. Also,
$p_-$ in (\ref{eq:77a}) can be zero due to
$1=\delta\sqrt{1+g^2/\omega^2}$|indicating that the fluctuations of
work are strictly non-negative|while $\pm \omega\sin[gt/2]$ can still
assume negative values with non-zero probability.

\section{A lower bound for $\Upsilon$}

The factor $\Upsilon$ is defined by (26) of the main text.

To derive a lower bound for $\Upsilon$ we minimize it over $\nu_k$
under the constraint $\sum_k\nu_k=1$ using Lagrange multipliers
(recall that $\nu_k$ and $\mu_l$ are probabilities). This produces: 
\begin{eqnarray}
  \label{eq:140}
  \Upsilon \geq 
 \left[{\sum}_k\, p^{1/2}_k\,
  \left({\sum}_l\,
 p_{k|l}\mu_l \right)^{1/2}\right]^2.
\end{eqnarray}
This lower bound is achievable and its RHS is smaller than $1$,
because it is a squared overlap of two probability vectors: $p_k$ and
${\sum}_l\,
 p_{k|l}\mu_l$.  Hence $\ln \Upsilon$ in (26
) [of the main
text] can be negative.


\begin{thebibliography}{99}
\bibitem{foot}Similar drawbacks of the difference between two
  Heisenberg operators were noted in the context of 
  simultaneous quantum measurements \cite{ozawa_ozawa}.
\bibitem{foo2} Two recent papers \cite{wang,hu} studied fluctuations
  of work|by employing what effectively amount to the operator of work
  definition|time-locally along effective classical trajectories. For
  defining those trajectories they employed the notion of decoherent
  histories. Note that the drawback pointed out at (\ref{tut:}
) is
  just formulated on a consistent history for the Heisenberg operator
  of work.
\end{thebibliography}
\end{document}